\documentstyle[12pt]{article}

\def\beq{\begin{equation}}
\def\eeq{\end{equation}}
\def\barr{\begin{eqnarray}}
\def\earr{\end{eqnarray}}
\def\dag{\dagger}
\def\b{\bigskip}
\def\tr{{\rm tr}}
\def\l{\left(}
\def\r{\right)}

\begin{document}

\title{Mass Degeneracies In Self-Dual Models\footnote{UCONN-94-9;
hep-th/9411124}}

\author{\normalsize{Gerald Dunne} \\
\normalsize{Department of Physics}\\
\normalsize{University of Connecticut}\\
\normalsize{Storrs, CT 06269 USA}\\
   \\
\normalsize{dunne@hep.phys.uconn.edu} \\}


\maketitle

\begin{abstract}
An algebraic restriction of the nonabelian self-dual Chern-Simons-Higgs systems
leads to coupled abelian models with interesting mass spectra. The vacua are
characterized by embeddings of $SU(2)$ into the gauge algebra, and in the
broken phases the gauge and real scalar masses coincide, reflecting the
relation of these self-dual models to $N=2$ SUSY. The masses themselves are
related to the exponents of the gauge algebra, and the self-duality equation is
a deformation of the classical Toda equations.
\end{abstract}

\b\b\b
The self-dual Chern-Simons systems \cite{hong,klee,kao,dunne} may be
characterized by the fact that the energy density possesses a Bogomol'nyi lower
bound which is saturated by solutions to first-order self-duality equations.
The special form of the self-dual potential may also be fixed by embedding this
bosonic system into a supersymmetric model and imposing the condition of $N=2$
SUSY \cite{clee}. Each of these characterizations is familiar from other
self-dual systems \cite{bog,hlousek}. In this Letter we investigate another
characterization of self-dual models, in terms of the spectra of massive
excitations in the various vacua, when the model is viewed as a spontaneous
symmetry breaking system. This is of particular interest for Chern-Simons
systems because the Higgs mechanism works in an unfamiliar manner \cite{deser}.
We find that in self-dual Chern-Simons theories associated with the
simply-laced A-D-E Lie algebras the gauge and real scalar mass spectra are
degenerate in each of the (many) inequivalent vacua. Furthermore, the masses
are given by a simple universal mass formula in terms of the exponents of the
algebra.

The nonabelian relativistic self-dual Chern-Simons system
\cite{hong,klee,kao,dunne} is described by the following Lagrange density (in
$2+1$ dimensional spacetime)
\beq
{\cal L}=-\tr\left(\left(D_\mu \phi\right)^\dag D^\mu \phi\right) -\kappa
\epsilon^{\mu \nu \rho} \tr\left(\partial_\mu A_\nu A_\rho +{2\over 3}
A_\mu A_\nu A_\rho \right) - V\left(\phi, \phi^\dag\right)
\label{lag}
\eeq
where the gauge invariant scalar field potential $V(\phi, \phi^\dag)$ is
\beq
V\l\phi, \phi^\dag\r = {1\over 4\kappa^2}\tr\left\{ \l[\;[\;\phi,
\phi^\dag\;],\phi\;]-v^2\phi\r^\dag\;\l[\;[\;\phi, \phi^\dag\;],\phi\;]-
v^2\phi\r\right\}.
\label{pot}
\eeq
The covariant derivative is $D_\mu \equiv\partial_\mu+[A_\mu,~\;]$, the
space-time metric is taken to be $g_{\mu\nu}={\rm diag}\l -1,1,1 \r$, and
$\tr$ refers to the trace in a finite dimensional representation of the
compact simple Lie algebra ${\cal G}$ to which the gauge fields $A_\mu$
and the charged scalar matter fields $\phi$ and $\phi^\dag$ belong. For the
particular sixth order potential (\ref{pot}) the energy density of this system
has a Bogomol'nyi lower bound which is saturated by solutions to the
 ``{\it relativistic self-dual Chern-Simons equations}'':
\barr
D_- \phi &=&0
\label{sd1}
\earr
\barr
F_{+-} &=& {1\over \kappa^2} [ v^2 \phi -[  [ \phi, \phi^\dag ], \phi]
,\phi^\dag  ]
\label{sd2}
\earr
where $D_- \equiv D_1-iD_2$, and $F_{+-}$ is the gauge curvature. The
nonrelativistic limit of this system is obtained by ignoring the quartic $\phi$
term in (\ref{sd2}), in which case we arrive at the nonrelativistic self-dual
Chern-Simons equations which are integrable \cite{djpt}. In the relativistic
case the general solution is not known, even in the abelian model \cite{hong},
although certain properties of the abelian solutions can be obtained from
asymptotic analysis and a radial ansatz. The only closed-form solutions known
in the relativistic case correspond to the zero energy solutions, for which the
gauge field is pure gauge and the $\phi$ field is gauge equivalent to a
solution of the {\it algebraic} equation:
\beq
[ [ \phi, \phi^\dag ], \phi ] =v^2 \phi .
\label{embedding}
\eeq
Solutions of this equation also correspond to the minima of the potential
(\ref{pot}), and these potential minima are clearly degenerate
\cite{kao,dunne}.

Here we propose to consider a simplified form of this self-dual system in which
the fields are algebraically restricted by the following ansatz:
\barr
\phi&=& \sum_{a=1}^{r}\phi^a E_a\cr
A_\mu &=& i \sum_{a=1}^{r} A_\mu^a h_a
\label{ansatz}
\earr
where $r$ is the rank of the gauge Lie algebra, and the $E_a$ and $h_a$ refer
to the simple root step operators and Cartan subalgebra generators
(respectively) in a ``Gell-Mann'' basis (for simplicity of notation, we
consider only the simply-laced algebras):
\barr
[h_a , h_b ]=0\hskip .5in [E_a, E_{-b} ] &=&\delta_{a\, b}\sum_{c=1}^r
\alpha^{(a)}_c h_c \hskip .5in [h_a , E_{b}]=\alpha^{(b)}_a E_{b}\cr
tr\left( E_a E_{-b} \right)=\delta_{ab} \hskip .5in tr\left( h_a h_b \right)
&=& \delta_{ab}\hskip 1in tr\left( h_a E_b \right) =0
\label{chevalley}
\earr
Here $K_{ab}$ is the Cartan matrix which encodes the inner products of the
simple roots $\vec{\alpha}^{(a)}$ (which have been normalized to each have
${\rm (length)}^2=2$):
\barr
K_{a\, b} \equiv
\vec{\alpha}^{(a)}\cdot\vec{\alpha}^{(b)}
\label{cartan}
\earr
With this algebraic restriction on the fields the original Lagrange density
(\ref{lag}) and potential (\ref{pot}) simplify considerably. Since the gauge
fields lie in the Cartan subalgebra, the Chern-Simons term in the Lagrange
density decomposes into $r$ copies of an abelian Chern-Simons term, leading to
\barr
{\cal L}_{\rm restricted} =- \sum_{a=1}^r \left| \partial_\mu \phi^a +
i\left(\sum_{b=1}^r A^b_\mu \alpha^{(a)}_b \right)\phi^a\right|^2 -\kappa
\sum_{a=1}^r \epsilon^{\mu\nu\rho} \partial_\mu A_\nu ^a A_\rho^a-V
\label{restrictedlag}
\earr
where the potential (\ref{pot}) becomes
\barr
V_{\rm restricted}={v^4\over 4\kappa^2} \sum_{a=1}^r | \phi^a |^2 -{v^2 \over
2\kappa^2}\sum_{a=1}^r\sum_{b=1}^r|\phi^a|^2 K_{ab} |\phi^b|^2 +{1\over
4\kappa^2}\sum_{a=1}^r\sum_{b=1}^r\sum_{c=1}^r |\phi^a|^2 K_{ab} |\phi^b|^2
K_{bc}|\phi^c|^2
\label{restrictedpot}
\earr
The Lagrange density (\ref{restrictedlag}) describes $r$ abelian Chern-Simons
gauge fields $A_\mu^a$ coupled to $r$ complex scalar fields $\phi^a$, with the
couplings determined by the Cartan matrix of the original nonabelian gauge
algebra. With this algebraic restriction, the relativistic self-dual
Chern-Simons equations (\ref{sd1},\ref{sd2}) combine into the single set of
coupled equations:
\barr
\partial_+ \partial_- ln |\phi^a|^2= -{v^2\over \kappa^2} \sum_{b=1}^r
K_{ab}|\phi^b|^2+{1\over \kappa^2}\sum_{b=1}^r\sum_{c=1}^r
|\phi^b|^2K_{bc}|\phi^c|^2K_{ac}
\label{deformation}
\earr
In the nonrelativistic limit (with factors of the speed of light $c$ restored
\cite{hong,klee}), the quartic term on the RHS of (\ref{deformation}) drops out
and one is left with the classical Toda system, which is known to be integrable
and explicitly solvable in terms of $r$ holomorphic functions. In the
relativistic case, the self-duality equation (\ref{deformation}) is a {\it
deformation} of the Toda system.

In this Letter we treat the restricted self-dual model with Lagrange density
(\ref{restrictedlag}) as a spontaneous symmetry breaking problem, with the
gauge field acquiring masses by the Chern-Simons-Higgs mechanism. The potential
has a number of (degenerate) minima, and we shall examine the spectrum of
fluctuations about these various vacua.

To investigate the vacuum structure we first identify the zero energy solutions
$\phi_{(0)}$. These correspond to the minima of the potential and so correspond
to solutions of (\ref{embedding}) with the field $\phi$ restricted as in
(\ref{ansatz}). With this ansatz, the condition (\ref{embedding}) becomes
\barr
\sum_{a=1}^{r}|\phi_{(0)}^a |^2 \phi_{(0)}^b K_{b a}=v^2\phi_{(0)}^b
\label{condition}
\earr
When $\phi_{(0)}^a \neq 0$ for all $a$, the solution is given by
\barr
|\phi_{(0)}^a|^2 = \sum_{b=1}^r \l K^{-1}\r_{ab}
\label{maximal}
\earr
where $K^{-1}$ is the inverse Cartan matrix. This can also be characterized as
follows: in the representation theory of Lie algebras \cite{humphreys}, an
important role is played by the sum $\vec{\rho}$ of all fundamental weights
$\vec{\lambda}^{(a)}$, which is also equal to one half times the sum of all
positive roots. This sum may be re-expanded in terms of the simple roots
$\vec{\alpha}^{(a)}$, and when this is done, the coefficients are equal to one
half times the values for $(\phi_{(0)}^a)^2$ given in (\ref{maximal}).
\barr
\vec{\rho}=\sum_{a=1}^r \vec{\lambda}^{(a)} = {1\over 2}\sum_{\alpha >0}
\vec{\alpha} = {1\over 2}\sum_{a=1}^r (\phi_{(0)}^a)^2 \vec{\alpha}^{(a)}
\label{weightsum}
\earr
Table 1 lists the vector $\vec{\rho}$, expanded in the basis of simple roots
$\vec{\alpha}^{(a)}$, for each of the simply-laced algebras. We call this
solution (\ref{maximal}) the ``principal embedding'' vacuum, for a reason to be
explained below.

\begin{table}[h]
\center
\begin{tabular}{|c|cccccccc|} \hline
\multicolumn{1}{|c|}{\rm Algebra}&\multicolumn{8}{c|}{\rm Coefficients
$(\phi_{(0)}^a)^2$ in $\vec{\rho}=(1/2)\sum_{a}^r (\phi_{(0)}^a)^2
\vec{\alpha}^{(a)}$} \\
\hline
$A_r$&r&2(r-1)&3(r-2)&\dots&3(r-2)&2(r-1)&r& \\ \hline
$D_r$&2r-2&2(2r-3)&3(2r-4)&\dots&(r-2)(r+1)&r(r-1)/2&r(r-1)/2 &\\ \hline
$E_6$&16&30&42&30&16&22&&\\ \hline
$E_7$&34&66&96&75&52&27&49&\\ \hline
$E_8$&58&114&168&220&270&182&92&136 \\ \hline
\end{tabular} \\
\caption{The expansion of $\vec{\rho}$, half the sum of positive roots, in the
basis of simple roots. The coefficients $(\phi_{(0)}^a)^2$ give the components
of the ``principal embedding'' vacuum solution (\protect\ref{maximal}).}
\label{sun}
\end{table}

Alternatively, one can look for solutions to (\ref{condition}) for which one or
more of the coefficients $\phi_{(0)}^a$ is zero. If, for example,
$\phi_{(0)}^b=0$, then equation (\ref{condition}) decouples into two or more
equations for the remaining nonzero coefficients. Effectively this amounts to
deleting the $b^{th}$ row and $b^{th}$ column from the Cartan matrix $K$ and
solving the decoupled conditions. The deletion of the $b^{th}$ row and $b^{th}$
column from the Cartan matrix can be represented diagrammatically as the
deletion of the $b^{th}$ dot from the Dynkin diagram of ${\cal
G}$.\footnote{This is because the Cartan matrix $K$ may be viewed as the
connection matrix for the Dynkin diagram in the sense that an off-diagonal
entry $K_{ab}=-1$ means that the $a^{th}$ and $b^{th}$ dots in the Dynkin
diagram are connected by a single line.} This deletion divides the Dynkin
diagram into two or more Dynkin diagrams for algebras of smaller rank. The
general solution to the vacuum condition (\ref{condition}) consists of taking
the principal embedding solution (\ref{maximal}) for each of these subdiagrams.
In $A_r\equiv SU(r+1)$ there are $p(r+1)$ ways of performing these successive
deletions, where $p(r+1)$ is the number of (unrestricted) partitions of $r+1$.
This gives the total number of inequivalent vacua for the $A_r$ system
\cite{dunne}.

We now determine the spectrum of massive excitations in these various vacua. In
the {\it unbroken} vacuum, with $\phi_{(0)}=0$, there are $r$ complex scalar
fields, each with mass
\barr
m={v^2\over 2\kappa}
\label{mass}
\earr
In one of the broken vacua, with $\phi_{(0)}\neq 0$, some of these $2r$ real
massive scalar degrees of freedom are converted to massive gauge degrees of
freedom. We shall concentrate initially on the principal embedding vacuum
(\ref{maximal}). This vacuum has the property that {\it all} $r$ gauge modes
acquire a mass, leaving $r$ massive (real) scalar modes. The new scalar masses
are determined by expanding the shifted potential $V(\phi+\phi_{(0)})$ to
quadratic order in the field $\phi$. For the principal embedding vacuum
(\ref{maximal}) this leads to:
\barr
V(\phi +\phi_{(0)})={v^4\over \kappa^2}\tr\l \left | [\phi_{(0)},
[\phi_{(0)}^\dag ,\phi ] \right |^2\r
={v^4\over \kappa^2}\sum_{a=1}^r\sum_{b=1}^r \phi^a \left( \phi_{(0)}^a
\phi_{(0)}^b \sum_{c=1}^r (\phi_{(0)}^c)^2 K_{ac}K_{bc} \right) \phi^b
\label{scalarmasses}
\earr
The real scalar masses are then given by the square roots of the eigenvalues of
the $r\times r$ mass (squared) matrix:
\barr
{\cal M}_{a b}^{\rm (scalar)}= 4\;m^2\; \phi_{(0)}^a \phi_{(0)}^b \sum_{c=1}^r
(\phi_{(0)}^c )^2 \, K_{a c}\, K_{b c}
\label{scalarmatrix}
\earr
where the $\phi_{(0)}^a$ are given by (\ref{maximal}). The resulting mass
spectra are listed in Table \ref{scalar} for the simply-laced algebras. Note
that all the masses are {\it integer} multiples of the mass scale $m$ in
(\ref{mass}).

\begin{table}[t]
\center
\begin{tabular}{|c|cccccccc|} \hline
\multicolumn{1}{|c|}{\rm Algebra}&\multicolumn{8}{c|}{\rm Masses}\\ \hline
$A_r$&2&6&12&20&30&\dots&r(r-1)&r(r+1) \\ \hline
$D_r$&2&12&30&56&90&\dots&(2r-3)(2r-2)&r(r-1)\\ \hline
$E_6$&2&20&30&56&72&132&&\\ \hline
$E_7$&2&30&56&90&132&182&306&\\ \hline
$E_8$&2&56&132&182&306&380&552&870 \\ \hline
\end{tabular} \\
\caption{The scalar masses, in units of $m$, for the principal embedding vacuum
(\protect\ref{maximal}), obtained as square roots of the eigenvalues of the
mass matrix in (\protect\ref{scalarmatrix}).}
\label{scalar}
\end{table}

The gauge masses are determined by expanding $\tr\l \left|
D_\mu\l\phi+\phi_{(0)}\r\right| ^2\r$ and extracting the piece quadratic in the
gauge field $A$:
\barr
v^2 \; \tr\l [A_\mu ,\phi_{(0)}]^\dag [A^\mu , \phi_{(0)} ]\r
\label{gaugemasses}
\earr
In the principal embedding vacuum, this leads to the following $r\times r$ mass
matrix:
\barr
{\cal M}_{a b}^{\rm (gauge)} =  2\;m\;\sum_{c=1}^r (\phi_{(0)}^c )^2\,
\alpha_a^{(c)}\,\alpha_b^{(c)}
\label{gaugemass}
\earr
where the $\phi_{(0)}^c$ are given by (\ref{maximal}).

Here we stress an important difference between the Chern-Simons-Higgs mechanism
and the conventional Higgs mechanism in a Yang-Mills gauge theory. In a
Yang-Mills theory, the gauge masses produced by the Higgs mechanism would be
the {\it square roots} of the eigenvalues of the mass matrix obtained from
(\ref{gaugemasses}). However, for gauge masses produced by the Chern-Simons
Higgs mechanism \cite{deser}, the masses are just the eigenvalues of the mass
matrix. This difference is essentially because the Chern-Simons Lagrange
density is {\it first order} in spacetime derivatives. By explicit computation,
the eigenvalues of the gauge mass matrix (\ref{gaugemass}) yield a gauge mass
spectrum which coincides exactly with the scalar mass spectrum presented in
Table \ref{scalar}. In other words, the eigenvalues of (\ref{scalarmatrix}) are
the squares of the eigenvalues of (\ref{gaugemass}).

Thus, the mass spectra of the gauge and scalar modes are {\it degenerate} in
the principal embedding vacuum. Such a degeneracy had been noted in the abelian
system, which corresponds to the $A_1\equiv SU(2)$ case of the present model
(\ref{restrictedlag},\ref{restrictedpot}). In the abelian model \cite{hong}
there is only one nontrivial vacuum, and  a consequence of the particular
$6^{th}$ order self-dual form of the potential is that in this broken vacuum
the massive gauge excitation and the real massive scalar field are degenerate
in mass. This was also found to be true for the $SU(N)$ system \cite{dunne}.
This pairing of the masses is a reflection of the fact that the self-dual
Chern-Simons system (\ref{lag},\ref{pot}) (and hence also the system
(\ref{restrictedlag},\ref{restrictedpot})) can be embedded into an $N=2$
supersymmetric model \cite{clee}. Such an embedding would be impossible if the
bosonic masses did not pair up in each vacuum. The $2+1$ dimensional Abelian
Higgs model, which can also be embedded into an $N=2$ SUSY theory at the
self-dual point, also has the feature that, with the self-dual potential, the
gauge and scalar masses in the broken vacuum are degenerate \cite{bog}.

To explain the algebraic origin of this remarkable mass degeneracy of the
system (\ref{restrictedlag},\ref{restrictedpot}), and to explain the particular
masses that arise, we reconsider the vacuum condition (\ref{embedding}). With a
factor of $v$ absorbed into the fields, this can be viewed as an embedding of
$SU(2)$ into the original gauge algebra ${\cal G}$.\footnote{It is interesting
to note that this type of embedding problem also plays a significant role in
the theory of instantons and of spherically symmetric magnetic monopoles and
the Toda molecule equations \cite{leznov}.} Thus, the vacuum solution
$\phi_{(0)}$ may be identified with an $SU(2)$ raising operator $J_+$, and so
on. Then the quadratic gauge field term in (\ref{gaugemasses}) may be re-cast
in terms of the adjoint action of $SU(2)$ on the gauge algebra ${\cal G}$:
\barr
m\, \tr \l A_\mu \l J_+ J_- + J_- J_+ \r A^\mu \r=m\, \tr \l A_\mu \l {\cal C}
- J_3^2 \r A^\mu \r
\label{adjointgauge}
\earr
where ${\cal C}$ is the $SU(2)$ quadratic Casimir. But $J_3 A^\mu =0$ since the
gauge fields are restricted to the Cartan subalgebra by the ansatz
(\ref{ansatz}). Thus the gauge masses are just given by the eigenvalues of the
quadratic Casimir ${\cal C}$ in the adjoint action (corresponding to the
particular $SU(2)$ embedding) of $SU(2)$ on the gauge algebra ${\cal G}$. It is
a classical result of Lie algebra representation theory \cite{dynkin} that the
adjoint action of the ``principal $SU(2)$ embedding'' (\ref{maximal}) on ${\cal
G}$ divides the $d\times d$ dimensional adjoint representation of ${\cal G}$
into $r$ irreducible sub-blocks, each of dimension $(2s_a+1)$ where the $s_a$
are known as the {\it exponents} of the algebra ${\cal G}$. Here $d$ is the
dimension of the algebra ${\cal G}$. This sub-blocking fills out the entire
$d\times d$ adjoint representation since the exponents have the property that
\barr
\sum_{a=1}^r (2 s_a+1) = d
\label{dimension}
\earr
The elements of each irreducible sub-block are arranged according to their
corresponding {\it principal grading} which is their $J_3$ eigenvalue.
Restricting to the Cartan subalgebra (as is acieved by the ansatz
(\ref{ansatz}) for the gauge fields) selects the $j_3=0$ element from each
sub-block, and in each irreducible sub-block the quadratic Casimir ${\cal C}$
has eigenvalue ${\cal C} = s_a (s_a + 1)$. The exponents for the classical
simply-laced Lie algebras are listed in Table \ref{exp}. It is straightforward
to verify that the mass spectrum in Table \ref{scalar} for the eigenvalues of
the gauge mass matrix (\ref{gaugemass}) coincides with the general mass formula
\barr
m_a = m\, s_a(s_a+1) \hskip 1in a=1, \dots r
\label{masses}
\earr
where the $s_a$ are the {\it exponents} of ${\cal G}$.

\begin{table}[t]
\center
\begin{tabular}{|c|c|c|cccccccc|} \hline
\multicolumn{1}{|c|}{\rm Algebra}&\multicolumn{1}{c|}{\rm
Rank}&\multicolumn{1}{c|}{\rm Dimension}&\multicolumn{8}{c|}{\rm Exponents} \\
\hline
$A_r$&r&r(r+2)&1&2&3&\dots&r-1&r&& \\ \hline
$D_r$&r&r(2r-1)&1&3&5&\dots&2r-3&r-1&&\\ \hline
$E_6$&6&78&1&4&5&7&8&11&&\\ \hline
$E_7$&7&133&1&5&7&9&11&13&17&\\ \hline
$E_8$&8&248&1&7&11&13&17&19&23&29 \\ \hline
\end{tabular} \\
\caption{The ranks, dimensions and exponents of the simply-laced classical Lie
algebras. Note that the sum of the exponents equals the number of positive
roots, which is one half (dimension -- rank).}
\label{exp}
\end{table}
To see that the real scalar masses are also given by the general mass formula
(\ref{masses}), we note that the quadratic part (\ref{scalarmasses}) of the
shifted scalar potential can be written as
\barr
4 m^2 \tr \l \phi^\dag \l J_+ J_- \r^2 \phi \r = m^2 \tr \l \phi^\dag \l {\cal
C} -J_3^2+J_3\r^2 \phi\r
\label{adjointscalar}
\earr
But $J_3\phi=1\phi$ since $\phi$ is expanded in terms of the simple root step
operators (and hence has principal grading 1). Thus the eigenvalues of the
scalar $mass^2$ matrix are the squares of the eigenvalues of ${\cal C}$, and we
find a scalar mass spectrum identical with the gauge mass spectrum in
(\ref{masses}).

For any vacuum $\phi_{(0)}$ {\it other} than the principal embedding one
(\ref{maximal}), the gauge and scalar masses may be found as follows. If the
vacuum solution $\phi_{(0)}$ corresponds to $n$ deletions of dots from the
original Dynkin diagram (as described before - see also \cite{dunne}) then $n$
complex scalar fields remain massive with mass $m$ corresponding to the scalar
mass in the unbroken vacuum. The remaining $(r-n)$ real scalar masses are
obtained from formula (\ref{masses}) using the exponents for each of the Dynkin
sub-diagrams. This also yields the $(r-n)$ real gauge masses. Thus in any
vacuum, the masses are always paired, either because they correspond to a
complex scalar field (of which the extreme case is the unbroken vacuum) or
because the real scalar and gauge masses coincide through formula
(\ref{masses}) (of which the principal embedding vacuum (\ref{maximal}) is the
extreme case).

To conclude, we have shown that in the coupled self-dual Chern-Simons systems
(\ref{restrictedlag},\ref{restrictedpot}) the gauge and real scalar masses in
the broken vacua are degenerate and are given by the universal mass formula
(\ref{masses}) in terms of the exponents of the associated Lie algebra ${\cal
G}$. It is interesting to note that in the affine Toda systems, the masses are
also related to the exponents of the associated gauge algebra \cite{freeman}.
The quantum implications, associated with tunnelling effects for example, of
these mass degeneracies and mass spectra remain to be resolved.

\vskip 1in

\noindent{\bf Acknowledgements:} This work has been supported in part by
the D.O.E. through grant number DE-FG02-92ER40716.00, and by the University
of Connecticut Research Foundation. I am grateful to Ed Corrigan for a helpful
suggestion.

\end{document}